\documentclass[twocolumn,aps,showpacs]{revtex4}

\usepackage{graphicx}
\usepackage{dcolumn}
\usepackage{bm}

\begin{document}

\preprint{APS/123-QED}

\title{Absolute frequency measurement of the \\ magnesium intercombination transition ${^1\!S_0}\rightarrow{^3\!P_1}$}

\author{Jan Friebe}
\email{friebe@iqo.uni-hannover.de}
\author{Andr\'e Pape}%
\author{Matthias Riedmann}
\author{Karsten Moldenhauer}
\author{Tanja {Mehlst\"aubler}}
\author{Nils Rehbein}
\author{Christian Lisdat}
\author{Ernst M. Rasel}
\author{Wolfgang Ertmer}
\affiliation{Institute of Quantum Optics, Leibniz {Universit\"at} Hannover, Welfengarten 1, 30167 Hannover, Germany}%

\author{Harald Schnatz}
\author{Burghard Lipphardt}
\author{Gesine Grosche}
\affiliation{Physikalisch-Technische Bundesanstalt, Bundesallee 100, 38116 Braunschweig, Germany}%

\date{\today}

\begin{abstract}
We report on a frequency measurement of the $(3s^2)^1\!S_0 \rightarrow {(3s3p)^3\!P_1}$ clock transition of $^{24}$Mg on a thermal atomic beam. The intercombination transition has been referenced to a portable primary Cs frequency standard with the help of a femtosecond fiber laser frequency comb. The achieved uncertainty is $2.5\times10^{-12}$ which corresponds to an increase in accuracy of six orders of magnitude compared to previous results. The measured frequency value permits the calculation of several other optical transitions from $^1\!S_0$ to the $^3\!P_J$-level system for $^{24}$Mg, $^{25}$Mg and $^{26}$Mg. We describe in detail the components of our optical frequency standard like the stabilized spectroscopy laser, the atomic beam apparatus used for Ramsey-Bord\'e interferometry and the frequency comb generator and discuss the uncertainty contributions to our measurement including the first and second order Doppler effect. An upper limit of $3\times10^{-13}$ in one second for the short term instability of our optical frequency standard was determined by comparison with a GPS disciplined quartz oscillator.
\end{abstract}

\maketitle

\section{Introduction}
\label{intro}

\begin{figure}
\resizebox{0.75\columnwidth}{!}{
  \includegraphics{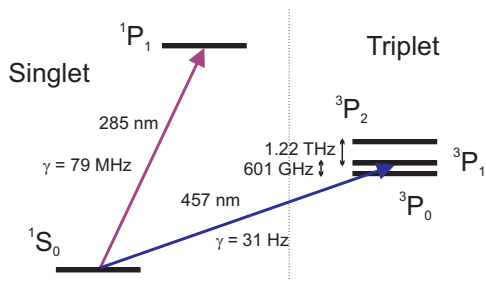}
}
\caption{(Color online) Level scheme for $^{24}$Mg displaying the typical features of neutral atom optical frequency standards like a strong cooling transition and a narrow intercombination line.}
\label{fig:levels}
\end{figure}

Neutral alkaline-earth metals belong to the small group of candidates \cite{Gill} for next generation optical atomic clocks. Systems using strontium and calcium \cite{Boyd2,Lemonde,KatoriJap,Ido,Degenhardt,wilpers} have recently been shown to outperform state-of-the-art Cs clocks in terms of stability and the accuracy of some optical ion frequency standards is already limited by the accuracy of the best Cs clocks \cite{Oskay,tamm}.

In this paper we present the first absolute frequency measurement of the $^1\!S_0 \rightarrow {^3\!P_1}$ intercombination line in magnesium. Our measurement which is based on atomic beam interferometry improves previous results obtained by means of grating spectrometry \cite{risberg} by six orders of magnitude. It can be used as an anchor for further measurements on cold free falling or trapped atoms. We use it to calculate other transitions in the $^1\!S_0 \rightarrow {^3\!P_J}$ system (table \ref{3P}) including the strictly forbidden $^1\!S_0 \rightarrow {^3\!P_0}$ transition.



A magnesium lattice clock, based on the transition $^1\!S_0 \rightarrow {^3\!P_0}$ can be implemented by using an optical lattice at the magic wavelength around 432 nm \cite{ovsiannikov} and by tailoring the linewidth with an external magnetic field \cite{taichenachev}. According to \cite{dereviankoblack} magnesium has the lowest black-body radiation shift for the $^1\!S_0 \rightarrow {^3\!P_0}$ transition of all alkaline-earth metals which is e.g. 10 times smaller than in Sr. In future optical lattice clocks the knowledge of the black body radiation shift may limit the achievable accuracy.

This paper is organized as follows: Section \ref{setup} describes the experimental setup including the stabilized diode laser, the atomic beam apparatus and the fiber laser frequency comb. In section \ref{results} we present the results of the absolute frequency measurement of the magnesium intercombination transition as well as an investigation of the stability of this frequency standard. Finally, we discuss in detail the main contributions to the uncertainty of our frequency measurement in section \ref{uncertainties}.

\section{Experimental Setup}
\label{setup}

\begin{figure*} 
  \includegraphics[width=0.95\textwidth]{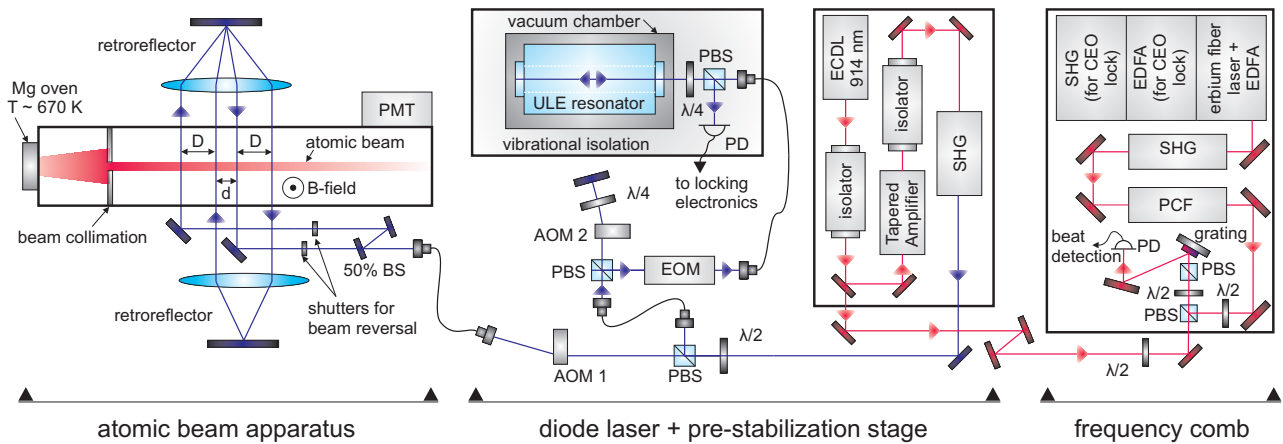}
\caption{(Color online) Setup of the magnesium optical frequency standard consisting of a stabilized diode laser system, an atomic beam apparatus and a commercial femtosecond fiber laser. EOM, electro-optic modulator; PD, photodetector; PBS, polarized beam splitter; BS, beam splitter; PCF, photonic-crystal fiber}
\label{fig:setup}
\end{figure*}

\subsection{Stabilized diode laser}
\label{laser}
\begin{figure}
\resizebox{1\columnwidth}{!}{
  \includegraphics{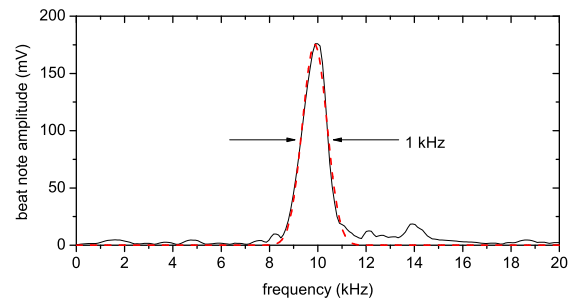}
}
\caption{(Color online) Beat note measurement at 457 nm between the stabilized diode and dye laser with a linewidth of about 870 Hz. RBW: 1 kHz, sweeptime 45 ms. Red dashed line: Gauss fit with FWHM of 1 kHz}
\label{fig:beat}
\end{figure}

The laser system used for interrogation of the magnesium clock transition is based on the commercially available TA-SHG 110 (Toptica Photonics). It consists of an extended cavity diode laser (ECDL) in Littrow configuration, a tapered amplifier and a resonant second harmonic generation (SHG) in bow-tie configuration with a KNbO$_3$ crystal. A small amount of the diode laser's light is picked up behind a deflection mirror and sent to the optical frequency comb for measurement purposes. Typically, this system delivers an output power of about 220 mW at 457 nm. The laser is locked to a 140 mm long high finesse ($\mathcal{F}=39\,000$) cavity made of ultralow-expansion material (ULE). This is mounted in a vacuum chamber on a granite block suspended from the ceiling for efficient suppression of acoustical and mechanical noise. The maximum observed drift of the laser frequency over several minutes is $\pm$3 Hz/s. Further details of our ULE cavity setup can be found elsewhere \cite{Keupp}. The error signal for the stabilization is obtained by the Pound-Drever-Hall technique (PDH) \cite{PDH}: A small amount of the blue light is transferred via a polarization maintaining single mode (pm) fiber to the suspended cavity platform as shown in fig. \ref{fig:setup} where the laser light is phase modulated by an EOM operated at a frequency of 9.7 MHz. An additional double-pass acousto-optic modulator (AOM) allows for frequency tuning. The PDH stabilization is accomplished via feedback to the laser diode current and the laser cavity piezoelectric transducer (PZT). To evaluate the linewidth of our laser system we performed a beat-note measurement with a dye laser  with a linewidth of about 870 Hz after stabilization to a Zerodur cavity \cite{Keupp}. Figure \ref{fig:beat} shows a beat-note linewidth of 1 kHz which gives an upper limit for the linewidth of our spectroscopy laser at the second harmonic wavelength.

\subsection{Atomic beam apparatus}
\label{atomic beam}
The magnesium clock is based on a thermal atomic beam Ramsey-Bord\'e apparatus. The collimated atomic beam has an average velocity of about 900 m/s. In our case the interferometer is realized with two pairs of antiparallel travelling light waves perpendicular to the atomic beam \cite{Borde} as shown in figure \ref{fig:setup}. The interaction region is enclosed by magnetic shielding. Small angular mismatch between the four beams leads to a reduction of contrast and introduces phase shifts in the interference pattern. Parallel preparation of all four laser beams is realized by means of two cat's eye retroreflectors \cite{snyder} each consisting of a mirror and a lens separated by the focal length of f=400 mm. A pm fiber delivers the light to the beam apparatus and acts as a spatial mode filter that generates a nearly perfect Gaussian beam profile. We have applied a constant magnetic field of $B = 6.77 \times 10^{-5} $ T to lift the degeneracy of the Zeeman sublevels.

\begin{figure}
\resizebox{0.95\columnwidth}{!}{
  \includegraphics{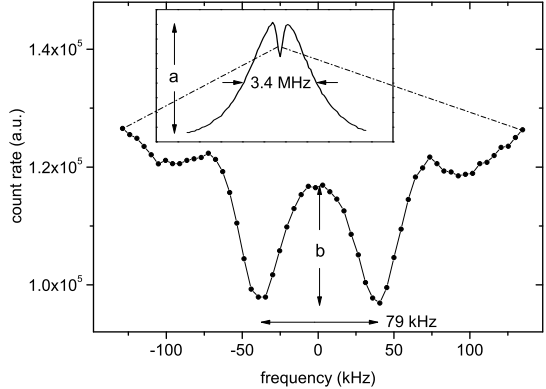}
}
\caption{Typical interference pattern observed with our apparatus. The averaging time per point is 2 s. The inset shows the complete spectrum on a larger frequency scale. We obtained a contrast of the interference pattern of b/a=12 \% which was limited by the maximum optical power of about 30 mW available at the experiment. }
\label{fig:ramseyfringe}
\end{figure}

The resolution of such an interferometer is only determined by the propagation time of the atoms between the co-propagating laser beams. The separation of the laser beams is adjusted to $D\cong4.5$ mm such that the signal of the two atom interferometers simultaneously formed in the Ramsey-Bord\'e scheme add up constructively. Figure \ref{fig:ramseyfringe} shows an interference pattern obtained with our apparatus by scanning the frequency of the laser across the resonance. The atoms excited in the interferometer to the state $\left|^3\!P_1, m_J=0\right>$ are measured by detection of the fluorescence of the atoms decaying to the ground state with a photomultiplier tube (PMT). Further details of this apparatus can be found elsewhere \cite{hinderthur}.

Frequency stabilization of our spectroscopy laser to the central fringe of the resonance is realized by a computer controlled square-wave modulation of the laser frequency. The interferometer is operated with this technique alternatingly at both sides of the central fringe. The error signal is deduced by subtracting the different PMT counts at the two points from each other. The modulation is performed using an AOM operated around 80 MHz in negative first order of diffraction. The digitally processed error signal serves to correct the offset frequency between laser and ULE cavity generated by AOM 2. To avoid offset drifts the AOM driver is referenced to the primary frequency standard.

\subsection{Frequency comb}

\label{comb}
We use the optical frequency synthesizer FC1500 (Menlo Systems) to generate the optical frequency comb spectrum \cite{Kubina}. The laser source of the FC1500 is a passively mode-locked femtosecond fiber laser \cite{Tamura2} which operates at a center wavelength of approximately 1550 nm and has a repetition rate of 100 MHz. The laser resonator has a small free-space section with one end mirror mounted on a translation stage controlled by a stepper motor. This permits the repetition rate to be tuned over approximately 400 kHz, with finer adjustments being achieved using a PZT.

The output power from the mode-locked laser is split and fed to two independent erbium-doped fiber amplifiers (EDFAs). The output from the first EDFA is broadened using a nonlinear fiber to span the wavelength range from approximately 1000 nm to 2100 nm. This provides the octave-spanning spectrum required for detection of the carrier-envelope offset frequency $\nu_{ceo}$ using the self-referencing technique. The f:2f interferometer is set up in a collinear single-arm configuration and uses a periodically-poled lithium niobate (PPLN) crystal to frequency double 2100 nm radiation to 1050 nm. After the PPLN crystal an interference filter selects out a narrow band of the spectrum around 1050 nm, resulting in a beat signal with a typical signal-to-noise ratio of 40-45 dB. The offset frequency is stabilized by feedback to the pump laser diode current.

A second EDFA generates high power radiation which is frequency doubled using a PPLN crystal, generating a narrow-band frequency comb around 780 nm. This is subsequently broadened in a nonlinear fiber to generate a comb spanning the range 580-940 nm. This second comb output is used to observe heterodyne beat signals with the fundamental frequency of our spectroscopy laser at 914 nm.

\section{Spectroscopic results}
\label{results}
\begin{figure}
\resizebox{0.95\columnwidth}{!}{
  \includegraphics{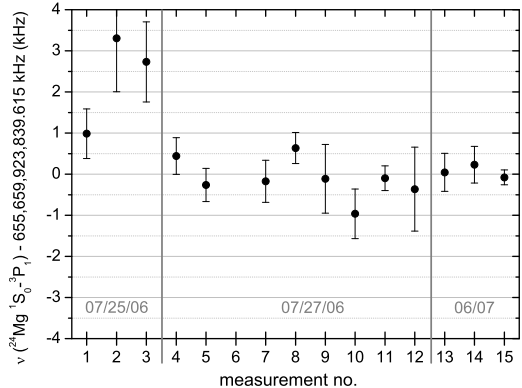}
}
\caption{Measurement results corrected for systematics with respect to the weighted average of all relevant values. Error bars only account for measurement uncertainties.}
\label{fig:measurement}
\end{figure}

We have determined the transition frequency of the magnesium intercombination line $(3s^2)^1\!S_0 \rightarrow {(3s3p)^3\!P_1}$ with respect to a portable Cs atomic clock (HP 5071A High Per\-for\-mance, accuracy: $\pm 2.0\times 10^{-13}$, stability: $\leq 5.0\times 10^{-12}$ in 1 s) from the Phy\-si\-ka\-lisch-Tech\-ni\-sche Bun\-des\-an\-stalt (PTB). Frequency measurements were performed in July 2006 on two different days and repeated in June 2007. Each time the interferometer was carefully realigned. The results are shown in figure \ref{fig:measurement}. Each data point is the arithmetic mean value of two frequency measurements, which were obtained by reversing the laser beams of the Ramsey-Bord\'e interferometer.
Notice that proper alignment of the beam reversal was not ensured in measurement no. 2 and 3 and these values have therefore been neglected for the determination of the average transition frequency.
Measurement no. 7 and 8 show the effect of reduced excitation intensity. In these experiments the optical power has been reduced to approx. 30 \% and 60 \% respectively. Since a reduced intensity leads to a change in excitation probability for different velocity classes it consequently directly affects the second order Doppler shift (see also sec. \ref{2nd Doppler}) and for this reason the values were also excluded and do not contribute to the average frequency value. To verify the effectiveness of our $\mu$-metal shielding the magnetic offset field has been switched off, inverted and returned to normal configuration in measurements no. 10-12. The measurements no. 13-15 were recorded approx. one year after the first measurement run and are in good agreement with the other frequency values.

The absolute frequency values in fig. \ref{fig:measurement} can be calculated according to:

\begin{equation}
\nu(^1\!S_0 \rightarrow {^3\!P_1})=2\,(2\,\nu_{ceo}+m\,f_{rep}+\nu_{fx})-80\,\rm MHz 
\label{frequenz aus beats}
\end{equation}

where $\nu_{fx} $ is the beat note between diode laser and femtosecond laser at 914 nm and the offset of -80 MHz is included due to AOM 1 (see fig. \ref{fig:setup}). The mode m of the fiber comb has been determined in two different ways:
As a first anchor for our target frequency we have calibrated our high precision wavelength meter (Highfinesse WS Ultimate, 40 MHz accuracy) with an iodine stabilized He-Ne-laser and shortly afterwards measured the frequency of the magnesium intercombination line. In a second experiment we applied a method proposed by Holzwarth et al. \cite{Holzwarth_2}, where the mode number m was determined for different repetition rates $f_{rep}$ and fixed laser frequency by applying eq. (\ref{frequenz aus beats}). 

\begin{table}
\caption{Shifts and uncertainties of the frequency measurement}
\label{Error measurement}
\begin{ruledtabular}
\begin{tabular}{lrrr}
Contribution & Shift\footnote{$\nu_{shift}=\nu_{observed}-\nu_{magnesium}$}&Uncertainty& relative\\
& in Hz &  in Hz & uncertainty \\
\hline
line shift\footnote{discussed in sec. \ref{uncertainties}}& -1641.1 & 1612 & $ 2.5 \times 10^{-12} $  \\
Cs clock & 0  &  131 & $ 2.0\times 10^{-13}$\\
Statistical uncertainty& 0 & 120 & $1.8 \times 10^{-13}$\\
\hline
Total & -1641.1 & 1622 & $ 2.5 \times 10^{-12}$ \\
\end{tabular}
\end{ruledtabular}
\end{table}

Table \ref{Error measurement} summarizes the uncertainty contributions to the measurement of the magnesium intercombination line.
The weighted average of all considered measurements has been determined to be 
\begin{equation}
\nu(^{24}Mg\,^1\!S_0 \rightarrow {^3\!P_1})=655\,659\,923\,839.6\,(± 1.6)\,\rm kHz
\label{frequenz}
\end{equation}
corresponding to a wavenumber of approx. 21870.4609 cm$^{-1}$. This value is in good agreement with, to our knowledge, the most recent published value in \cite{risberg} where a wavenumber of 21870.464 (0.02) cm$^{-1}$ has been determined in a measurement based on grating spectrometry which dates back to 1965. Compared to this result we have decreased the measurement uncertainty of the magnesium intercombination line frequency by six orders of magnitude. In a previous experiment which was performed in 1938 Mei\ss ner \cite{meissner} obtained a wavenumber of 21870.484 cm$^{-1}$. This value deviates approx. 700 MHz from our result.
Our measured frequency can be used as an anchor to calculate the other optical transitions from $^1\!S_0$ into the $^3\!P_J$ level system with the help of the well known radio frequency transitions in the triplet system as shown in table \ref{3P}.

\begin{table*}
\caption{Selection of optical transitions from $^1\!S_0$ to the $^3\!P_J$ level system, calculated from experimental data}
\label{3P}
\begin{ruledtabular}
\begin{tabular}{llrl}
Isotope & upper level & shift (relative to $^{24}$Mg $ {^3\!P_1} $) & absolute transition frequency ($ {^1\!S_0} \rightarrow {^3\!P_J} $) \\
\hline
$^{24}$Mg & $^3\!P_0$ & -601 277 157 869.1 (0.6) Hz \cite{Godone} & 655 058 646 681.7 (1.6) kHz  \\
$^{24}$Mg & $^3\!P_2$ & 1 220 575.1 (33) MHz \cite{Inguscio} & 656 880 498.9 (33) MHz \\
$^{25}$Mg & $^3\!P_1$ (F = 7/2) & 1040.7 (0.1) MHz \cite{Sterr} & 655 660 964.5 (0.1) MHz  \\
$^{25}$Mg & $^3\!P_1$ (F = 5/2) & 1556.8 (0.1) MHz \cite{Sterr} & 655 661 480.6 (0.1) MHz \\
$^{25}$Mg & $^3\!P_1$ (F = 3/2) & 1906.8 (0.1) MHz \cite{Sterr} & 655 661 830.6 (0.1) MHz  \\
$^{26}$Mg & $^3\!P_1$ & 2683.18 (0.02) MHz \cite{Sterr} & 655 662 607.02 (0.02) MHz \\
\end{tabular}
\end{ruledtabular}
\end{table*}

\begin{figure}
\resizebox{1\columnwidth}{!}{
  \includegraphics{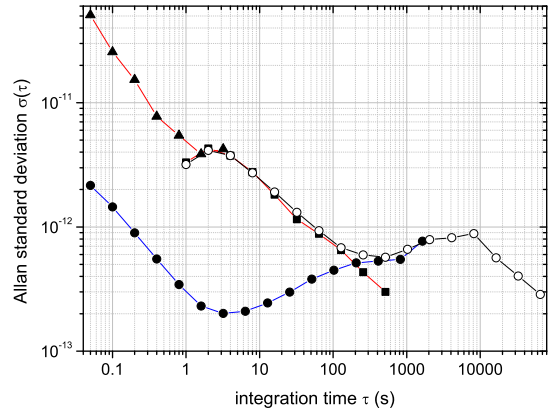}
}
\caption{(Color online) Allan standard deviation of our frequency standard with respect to a quartz oscillator (full circles, blue) and with respect to a portable Cs clock (full squares and triangles, red). Open circles: Cs clock vs. GPS disciplined quartz.}
\label{fig:allan}
\end{figure}

The stability of our frequency standard has been characterized with respect to a GPS disciplined quartz oscillator (Oscilloquartz OCXO 8607 Option 20) and the portable Cs atomic clock. The result is shown in fig. \ref{fig:allan}. The instability of the magnesium frequency standard and the quartz is about $3\times10^{-13}$ in one second. The instability of the quartz is specified to $2\times10^{-13}$ in one second. 
At integration times less than 100 s the stability measurement against the quartz oscillator sets an upper limit for the instability of the magnesium frequency standard.
All absolute frequency measurements have been performed with respect to the Cs clock since the GPS disciplined quartz reaches an accuracy better than $10^{-12}$ only for averaging times longer than 10,000 s.

The stability of our Ramsey-Bord\'e interferometer is investigated with the help of our pre-stabilized clock laser. For this purpose we recorded an interference pattern (fig. \ref{fig:ramseyfringe}) and subtracted a parabolic curve to remove the incoherent background. Subsequently a cosine term with exponential decreasing amplitude and quadratic increasing period was fitted to the resulting signal and subtracted. We calculated the Allan deviation $ \sigma^{(I)} $ of the residual differences and estimated the Allan deviation $ \sigma^{(\nu)} $ according to \cite{Riehle} via
\begin{equation}
\sigma^{(\nu)} = \frac{1}{\partial_\nu I} \sigma^{(I)}
\label{allandev}
\end{equation}
with $ \partial_\nu I $ denoting the first derivative of the Ramsey structure taken at the point used for stabilization.
Thus, we obtained a relative Allan standard deviation of $\sigma_y(1 {\;\rm s}) = 6 \times 10^{-13}$. This result is larger than the measured Allan standard deviation between quartz oscillator and the diode laser which is locked to the Ramsey-Bord\'e interferometer for longer integration times (see fig \ref{fig:allan}).

\section{Systematic uncertainties}
\label{uncertainties}

\subsection{Residual first order Doppler effect}
\label{1st Doppler}

Ideally Ramsey-Bord\'e spectroscopy strongly sup\-pres\-ses the first order Doppler effect. Small imperfections of the  interrogating beams or defocused retroreflectors can induce residual shifts and phase errors in the interference pattern. Moreover, due to unavoidable aberration in the retroreflector lenses perfect parallelism can only be obtained for three of the four beams. Studies in \cite{sengstock} showed that small deviations of a few microradians can lead to frequency shifts in the kHz range. Frequency shifts due to phase errors in the excitation zones can be identified for example by inverting the propagation direction of the excitating beam which results in a change of sign of the frequency shift induced by these effects. The arithmetic mean of the forward running beam measurement and the inverted beam measurement should reveal the correct transition frequency as has been pointed out by Kersten et al. \cite{Kersten}. 

The beam reversal is implemented with a 50 \% BS in front of the interaction zone. Both Ramsey-Bord\'e interferometers are carefully aligned such that the laser beams are perfectly overlapped and travel equal distances. The overlap of the laser beams for the two interferometer configurations with reversed laser beams was optimized along several meters at different points. 

The Ramsey-Bord\'e interferometry was optimized by individually adjusting the retroreflecting units. The first unit was aligned according to the procedure proposed in \cite{morinaga} with the help of saturation spectroscopy. The second retroreflector unit was aligned by optimising the contrast of the interference pattern.

\begin{figure}
\resizebox{0.95\columnwidth}{!}{
  \includegraphics{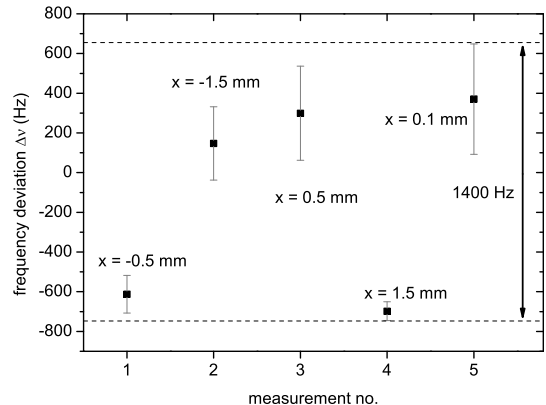}
}
\caption{Frequency deviation for different displacements (x) of one cat eye mirror from optimal position.}
\label{fig:beam reversal}
\end{figure}

We investigated if the beam reversal technique cancels systematic frequency shifts caused by misalignment of the second reflector unit. For this purpose we used our ULE resonator as a short term frequency reference. The result is shown in fig. \ref{fig:beam reversal} where a linear drift of the resonator has been subtracted. The mean frequency resulting from the beam reversal does not show any systematics with respect to the defocusing of the retroreflecting mirror. The maximum frequency difference of $\Delta\nu\approx1400$~Hz observed in this set of measurements was taken as a rough estimate for the upper limit of the error due to the residual Doppler effect also in the case of the optimized Ramsey-Bord\'e interferometer.

\subsection{Second order Doppler effect}
\label{2nd Doppler}

The Ramsey-Bord\'e interferometer cancels the Doppler effect only in first order of $ \beta $
. For thermal atomic beams with velocities of several hundred m/s, the quadratic term is of the order of a few kHz for transition frequencies in the optical domain, corresponding to a relative frequency uncertainty of $ 10^{-12} $. The estimation of the 2nd order Doppler effect follows \cite{Kersten}, where the experimentally observed interference pattern is compared with a calculated interference pattern without the term of the 2nd order Doppler effect. The experimental interference pattern can be expressed by:

\begin{eqnarray}
I(\Delta) = \int_0^\infty \! \! dv & g(v) & [ \cos \left( \frac{2D}{v} \left[ \Delta + \omega_0 \frac{v^2}{2 c^2} + \frac{\hbar k^2}{2 m} \right] \right) \nonumber \\
& + & \! \! \! \cos \left( \frac{2D}{v} \left[ \Delta + \omega_0 \frac{v^2}{2 c^2} - \frac{\hbar k^2}{2 m} \right] \right) ]
\label{Ramsey}
\end{eqnarray}

where the cosine terms represent the Ramsey signals of atoms passing the interaction zones (distance $ D $) with velocity $ v $. $ \Delta := \omega - \omega_0 $ is the detuning of the interrogating light with frequency $ \omega = c k $. $ g(v) $ is a velocity dependent weighting factor which accounts for thermal velocity distribution of the atoms and the complex dependencies of the excitation and detection probability on the atomic velocity. In order to calculate the interference signal without the 2nd order Doppler effect, the weighting factor $g(v)$ has to be determined. This can be done to sufficient approximation by Fourier transformation of ${I}(\Delta)$: 

\begin{eqnarray}
\tilde{I} (t) = \frac {D}{t^2} g \left( \frac{2D}{t} \right) \left( e^{it \frac{\hbar k^2}{2m}} + e^{-it \frac{\hbar k^2}{2m}} \right) \nonumber \\
\Leftrightarrow \; g \left( \frac{2D}{t} \right) = \frac{t^2}{2D \cos (t \frac{\hbar k^2}{2m})} \tilde{I}(t)
\label{distribution}
\end{eqnarray}

For the Fourier transform the second order Doppler effect is neglected. The modified interference signal without the 2nd order Doppler effect is given by

\begin{equation}
I(\Delta) = D \int_{-\infty}^\infty \! \! dt \; g \left( \frac{2D}{t} \right) \frac{1}{t^2} \left( e^{it \frac{\hbar k^2}{2m}} + e^{-it \frac{\hbar k^2}{2m}} \right) e^{it \Delta }
\label{FT eqn}
\end{equation}

where we performed the substitution $t=2D/v$.

\begin{figure}
\resizebox{\columnwidth}{!}{
  \includegraphics{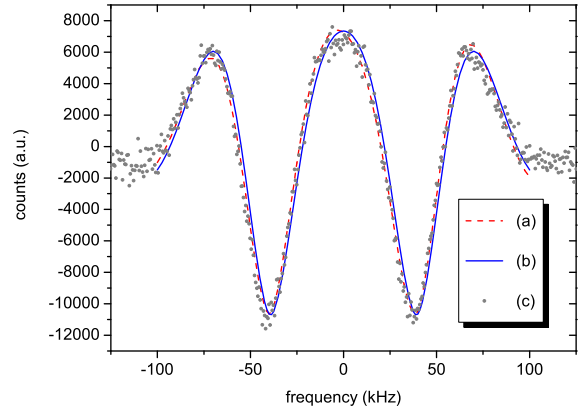}
}
\caption{(Color online) Theoretical Ramsey fringes with (a) and without second order Doppler shift (b) calculated from experimental parameters, (c) shows the experimental data points}
\label{Ramsey theo}
\end{figure}

We fitted a 2nd degree polynomial to the Lamb dip of our experimental data and subtracted it from the Ramsey fringes to eliminate the incoherent background. With the known distribution function $g(v)$ we were able to calculate the Ramsey fringes from equation (\ref{Ramsey}) for the case with and without 2nd order Doppler shift.
The quality of the approximation is tested by reinserting the evaluated velocity weighting factor in (\ref{Ramsey}) and by comparing the resulting signal with experimental data. As shown in fig. \ref{Ramsey theo}, the deviation is negligible compared to other uncertainties related to this effect like the knowledge of D. 
We calculated the whole fringe pattern as shown in fig. \ref{Ramsey theo} and determined due to our laser locking scheme (section \ref{atomic beam}) the shift on the sides of the central fringe and obtained a 2nd order Doppler shift of $ \Delta \nu = -1643 $ Hz.

In order to estimate the influence of the uncertainty of D we calculated the 2nd order Doppler effect for D=3.5~mm to 6 mm. From this we derived a maximum deviation of $ \pm 800 $ Hz. These results are in good agreement with the experimentally determined value of $ -1.5 \; (0.4) $ kHz by \cite{Sengstock_2}.

\subsection{External fields}
\label{fields}

The presence of a small static electric field inside the atomic beam apparatus (caused for example by terrestrial stray fields and non perfect conducting properties of the vacuum chamber materials) leads to a shift of atomic energy levels due to the Stark effect.
The difference of the polarizabilities of the $ ^1\!S_0 $ and $ ^3\!P_1 $ state were experimentally determined in \cite{Riegerdiss} as 
\mbox{$\Delta \alpha =\alpha(^3\!P_1,m=0) - \alpha(^1\!S_0) = 3.9 \; (0.1) \; \rm kHz \left(cm/kV\right)^{2}$}.
With $ \Delta \nu = -\frac{1}{2} \Delta \alpha E^2 $ and the assumption for the electrostatic rest field $ E $ to be $ 5 \; (5) $ V/cm we estimate a lineshift of $ \Delta \nu = - 0.05 \; (0.05) $ Hz.

The $ {^1\!S_0} \rightarrow {^3\!P_1} \; (m_J = 0) $ transition shows no first order Zeeman shift. However, this shift no longer vanishes in second or higher orders of B.
The correction term $ E_n^{(2)} $ of the n-th energy level $ E_n = \sum_{i=0}^{\infty} E_n^{(i)} $ in second order stationary perturbation theory has the form

\begin{equation}
E_n^{(2)} = \sum_{m \neq n} \frac{\left| \left< m^{(0)} \left|\hat{H}_I \right| n^{(0)} \right> \right|^2}{E_n^{(0)} - E_m^{(0)}}
\label{2nd perturbation}
\end{equation}

where $ \left| n^{(0)} \right> $, $ E_n^{(0)} $ is the unperturbed n-th state and energy level, respectively, and the sum runs over all unperturbed states. $ \hat{H}_I $ is the interaction Hamiltonian and in case of a static homogeneous magnetic field B (chosen to be in the direction of z-axis) given by

\begin{equation}
\hat{H}_I = \frac{e}{2 m_e} (\hat{L}_z + g_s \hat{S}_z)B_z
\label{H_I}
\end{equation}

$ \hat{L}_z $ and $ \hat{S}_z $ are the operators of angular momentum and electronic spin, $ e $ the elementary charge, $ m_e $ the electron mass and $ g_s \approx 2 $ the Land\'e factor. As the $ ^1\!S_0 $ state stays unshifted, we have to take into account only the $ ^3\!P_1 $ state. The correction terms are proportional to the inverse difference of the energy levels, so only the $ ^3\!P_0 $ and $ ^3\!P_2 $ states with $ \nu_{{^3\!P_1} \rightarrow {^3\!P_0}} = 601$ GHz and $ \nu_{{^3\!P_2} \rightarrow {^3\!P_1}} = 1.22 $ THz contribute significantly to the sum in (\ref{2nd perturbation}).  All other energy levels (with transition frequencies to $ ^3\!P_1 $ of more than a few hundred THz) can be neglected. By use of the Clebsch-Gordan coefficients we write the atomic $ \left| J, m_J \right> $ eigenstates as $ \left| L, m_L \right> \otimes \left| S, m_S \right> $ eigenstates of the $\hat{L}_z$, $\hat{S}_z$ operators, evaluate the sum in (\ref{2nd perturbation}) and arrive at:

\begin{equation}
E_{^3\!P_1}^{(2)} = \frac{1}{3} \left( \frac{e \hbar B}{2 m} \right)^2 \left[ \frac{2}{E_{^3\!P_1}^{(0)} - E_{^3\!P_0}^{(0)}} - \frac{1}{E_{^3\!P_2}^{(0)} - E_{^3\!P_1}^{(0)}} \right]
\label{2nd Zm shift}
\end{equation}

Therefore, the shift $ \Delta \nu $ of the transition frequency can be written after evaluating equation (\ref{2nd Zm shift}) with the numerical values as

\begin{equation}
\Delta \nu = 1.64 \times 10^8 \; {\rm \frac{Hz}{T^2} } \; B^2
\label{simple exp}
\end{equation}

By measuring the linear Zeeman shift of the $ ^3\!P_1\,(m_J=-1, m_J=1) $ substates we determined the strength of the applied homogeneous magnetic field to be $ B = 6.77 \; (0.45) \times 10^{-5} $ T. This causes a shift in second order of $ \Delta \nu = 0.75 \; (0.1) $ Hz far below the resolution of our apparatus.

\subsection{Sagnac effect}
\label{sagnac}

The thermal beam Ramsey-Bord\'e interferometer is sensitive to rotations. The resulting frequency shift can be calculated \cite{Riehle_sagnac} as

\begin{equation}
\Delta \nu = \frac{\Omega (D + d )}{\lambda}
\label{Sagnac_eq}
\end{equation}

with $ \lambda $ the wavelength of the interrogating light, $ D $ the distance between the co- and $ d $ between the counterpropagating beams (fig. \ref{fig:setup}). With values of $ D=4.5 \; (1.0)$~mm, $ d= 5.0 \; (1.0) $ mm and a rotation of the apparatus of $ \Omega= 5.75 \times 10^{-5} \; \rm s^{-1} $ which is caused by revolution of the earth  
we derive a frequency shift of $ \Delta \nu = 1.2 \; (0.1) $ Hz.

Table \ref{Error Mg} summarizes all relevant uncertainty contributions to the line shift of the magnesium frequency standard.

\begin{table}[htp]
\caption{Shifts and uncertainties of the $ ^{24}Mg\,\,(3s^2)^1\!S_0 \rightarrow {(3s3p)^3\!P_1}$ transition measured on a thermal atomic beam.}
\label{Error Mg} 
\begin{ruledtabular}
\begin{tabular}{lrrr}
Effect & Shift &Uncertain- & Relative\\
& in Hz & ty in Hz & uncertainty \\
\hline
1st order Doppler & 0 & 1400 & $ 2.1 \times 10^{-12} $  \\
2nd order Doppler & -1643 & 800 & $ 1.2 \times 10^{-12} $ \\
external magnetic field& 0.75 & 0.1 & $ 1.5\times 10^{-16}$\\
external electric field& -0.05 & 0.05 & $7.5\times 10^{-17}$\\
Sagnac effect& 1.2 & 0.1 & $1.5\times 10^{-16}$\\
\hline
Total & -1641.1 & 1612 & $ 2.5 \times 10^{-12}$ \\
\end{tabular}
\end{ruledtabular}
\end{table}

\section{Conclusion and Outlook}
\label{conclusion}

An optical frequency standard based on neutral magnesium has been realized. We determined the frequency of the intercombination line ${^1\!S_0}\rightarrow{^3\!P_1}$ at 457 nm with an accuracy of $2.5\times10^{-12}$. 
%
%
Our result deviates by 70 MHz from a value obtained by Risberg \cite{risberg} in 1965 which is a tenth of the quoted error and differs by 700 MHz from a previous measurement performed by Mei\ss ner in 1938. We used the measured frequency to calculate optical transitions from $^1\!S_0$ into the $^3\!P_J$ level system (table \ref{3P}). Especially the $^1\!S_0 \rightarrow {^3\!P_0}$ transition is interesting for future applications such as an optical lattice clock with magnesium.

By application of beam-reversal technique we were able to reduce one of the main uncertainty contributions in Ramsey-Bord\'e interferometry on thermal atoms, the residual first order Doppler effect, to a level of $2.1\times10^{-12}$. The residual linear and second order Doppler effect limit the achieved uncertainty of our frequency standard. A further considerable enhancement in accuracy can only be expected from measurements on laser cooled free falling or in a dipole trap captured atoms in combination with a primary frequency reference with superior stability compared to the presently used Cs clock. Currently we investigate possible experimental strategies \cite{malossi,mehlstaubler,binnewies,quench} to implement a lattice clock, which is expected to improve the current uncertainty in accuracy by several orders of magnitude.

\begin{acknowledgments}
This work was supported by the Deutsche Forschungsgemeinschaft (DFG) under SFB 407. The authors would like to thank Andreas Bauch from the PTB for providing a portable Cs atomic clock.
\end{acknowledgments}


\end{document}